\documentclass[prb,a4paper,twocolumn,floatfix,showpacs,showkeys,amsmath,amssymb,nobibnotes,altaffilletter,superscriptaddress]{revtex4-2}

\setcitestyle{super}
\usepackage{graphicx}
\usepackage[utf8]{inputenc}
\usepackage{amsmath}
\usepackage{xr}
\usepackage{upgreek}
\usepackage[caption=false]{subfig}
\usepackage{xspace}
\usepackage{textcomp} 
\usepackage{csquotes} 

\graphicspath{{.}{figures/}}
\newcommand*\sref[1]{S\ref{#1}} 
\newcommand{\degreeC}{\textdegree{}C\xspace}
\newcommand{\Jsc}{J_\mathrm{sc}}
\newcommand{\Voc}{V_\mathrm{oc}}
\newcommand{\FF}{F\!F}

\newcommand{\nid}{n_\mathrm{id}}
\newcommand{\eqepv}{\mathrm{EQE_{PV}}}
\newcommand{\el}{\mathrm{EL}}
\renewcommand{\d}{\ensuremath{\operatorname{d}\!}}

\begin{document}

\title{Traps and transport resistance: the next frontier for stable state-of-the-art non-fullerene acceptor solar cells}

\author{Christopher Wöpke}
\affiliation{Institut für Physik, Technische Universität Chemnitz, 09126 Chemnitz, Germany}

\author{Clemens Göhler}
\affiliation{Institut für Physik, Technische Universität Chemnitz, 09126 Chemnitz, Germany}

\author{Maria Saladina}
\affiliation{Institut für Physik, Technische Universität Chemnitz, 09126 Chemnitz, Germany}

\author{Xiaoyan Du}
\affiliation{Institute of Materials for Electronics and Energy Technology (i-MEET), Friedrich-Alexander-Universität Erlangen-Nürnberg, 91054 Erlangen, Germany}
\affiliation{Helmholtz Institute Erlangen-Nürnberg for Renewable Energy (HI ERN), Immerwahrstrasse 2, 91058 Erlangen, Germany}

\author{Li Nian}
\affiliation{Institute of Materials for Electronics and Energy Technology (i-MEET), Friedrich-Alexander-Universität Erlangen-Nürnberg, 91054 Erlangen, Germany}
\affiliation{Guangdong Provincial Key Laboratory of Optical Information Materials and Technology, Institute of Electronic Paper Displays, South China Academy of Advanced Optoelectronics, South China Normal University, Guangzhou, 510006 P. R. China}

\author{Christopher Greve}
\affiliation{Physikalisches Institut, Dynamik und Strukturbildung - Herzig Group, Universität Bayreuth, Universitätsstr. 30, 95447 Bayreuth, Germany}

\author{Chenhui Zhu}
\affiliation{Advanced Light Source, Lawrence Berkeley National Laboratory, Berkeley, California 94720, United States}

\author{Kaila M. Yallum}
\affiliation{Department of Chemistry and Biochemistry, University of Bern, 3012 Bern, Switzerland}

\author{Yvonne J. Hofstetter}
\affiliation{Integrated Center for Applied Photophysics and Photonic Materials and the Center for Advancing Electronics Dresden, Technische Universität Dresden, 01062 Dresden, Germany}

\author{David Becker-Koch}
\affiliation{Integrated Center for Applied Photophysics and Photonic Materials and the Center for Advancing Electronics Dresden, Technische Universität Dresden, 01062 Dresden, Germany}

\author{Ning Li}
\affiliation{Institute of Materials for Electronics and Energy Technology (i-MEET), Friedrich-Alexander-Universität Erlangen-Nürnberg, 91054 Erlangen, Germany}
\affiliation{Helmholtz Institute Erlangen-Nürnberg for Renewable Energy (HI ERN), Immerwahrstrasse 2, 91058 Erlangen, Germany}

\author{Thomas Heumüller}
\affiliation{Institute of Materials for Electronics and Energy Technology (i-MEET), Friedrich-Alexander-Universität Erlangen-Nürnberg, 91054 Erlangen, Germany}
\affiliation{Helmholtz Institute Erlangen-Nürnberg for Renewable Energy (HI ERN), Immerwahrstrasse 2, 91058 Erlangen, Germany}

\author{Ilya Milekhin}
\affiliation{Institut für Physik, Technische Universität Chemnitz, 09126 Chemnitz, Germany}

\author{Dietrich R.$\,$T. Zahn}
\affiliation{Institut für Physik, Technische Universität Chemnitz, 09126 Chemnitz, Germany}

\author{Christoph J. Brabec}
\affiliation{Institute of Materials for Electronics and Energy Technology (i-MEET), Friedrich-Alexander-Universität Erlangen-Nürnberg, 91054 Erlangen, Germany}
\affiliation{Helmholtz Institute Erlangen-Nürnberg for Renewable Energy (HI ERN), Immerwahrstrasse 2, 91058 Erlangen, Germany}

\author{Natalie Banerji}
\affiliation{Department of Chemistry and Biochemistry, University of Bern, 3012 Bern, Switzerland}

\author{Yana Vaynzof}
\affiliation{Integrated Center for Applied Photophysics and Photonic Materials and the Center for Advancing Electronics Dresden, Technische Universität Dresden, 01062 Dresden, Germany}

\author{Eva M. Herzig}
\affiliation{Physikalisches Institut, Dynamik und Strukturbildung - Herzig Group, Universität Bayreuth, Universitätsstr. 30, 95447 Bayreuth, Germany}

\author{Roderick C.$\,$I. MacKenzie}
\affiliation{Department of Engineering, Durham University, Lower Mount Joy, South Road, Durham, DH1 3LE, UK}

\author{Carsten Deibel}
\affiliation{Institut für Physik, Technische Universität Chemnitz, 09126 Chemnitz, Germany}
\email{deibel@physik.tu-chemnitz.de}

\begin{abstract}
Stability is one of the most important challenges facing organic solar cells (OSC) on their path to commercialization. In the high-performance material system PM6:Y6 studied here, investigate degradation mechanisms of inverted photovoltaic devices. We have identified two distinct degradation pathways: one requires presence of both illumination and oxygen and features a short-circuit current reduction, the other one is induced thermally and marked by severe losses of open-circuit voltage and fill factor. We focus our investigation on the thermally accelerated degradation. Our findings show that bulk material properties and interfaces remain remarkably stable, however, aging-induced defect state formation in the active layer remains the primary cause of thermal degradation. The increased trap density leads to higher non-radiative recombination, which limits open-circuit voltage and lowers charge carrier mobility in the photoactive layer. Furthermore, we find the trap-induced transport resistance to be the major reason for the drop in fill factor. Our results suggest that device lifetimes could be significantly increased by marginally suppressing trap formation, leading to a bright future for OSC.
\end{abstract}


\maketitle

\section{Introduction}

The power conversion efficiency of organic photovoltaic devices has increased from 2.5\% in 2001\cite{shaheen2001} to over 18\% today.\cite{Lin2020,Zhang2021} Despite this promising increase in performance, rooftop stability remains arguably the most pressing challenge. Device stability can be affected by many factors including material system,\cite{MatekerMcGehee} device architecture,\cite{Sherafatipour2019} deposition method,\cite{Riede_2008,hou2018review} morphology,\cite{Ramirez2021,Li2017} environmental conditions,\cite{MADOGNI2015201} and additives.\cite{LindqvistC}

Devices exposed to the atmosphere are vulnerable to processes involving oxygen and water. These degradation mechanisms can be accelerated in the presence of light. It has been widely reported that oxidation of interlayers and electrodes leads to the formation of extraction barriers for photogenerated charge carriers.\cite{Alam2020Disentanglement} Depending on the energetic landscape of the material system, photoinduced radicalization of molecular oxygen can lead to rapid degradation. Radicalized oxygen is highly reactive and has the potential to chemically react with most materials used within the device. Within the active layer such reactions can lead to bleaching, increased energetic disorder, and a reduction in photogeneration.\cite{Bregnhj2020Oxygen, Sudakov2020TheInterplay}

In an inert atmosphere the main degradation pathway in fullerene-based devices is the photo-induced dimerization of fullerene molecules, leading to a severely limited short-circuit current $\Jsc$ due to photobleaching.\cite{Heumueller2016Morphologial,Wang2021Recent,Ramirez2021} In order to avoid photo-induced dimerization, considerable attention has recently been focused on non-fullerene acceptor (NFA)-based devices. 
Dimerization also occurs in NFAs if the energetic alignment between donor and acceptor allows for back-hole transfer, back-electron transfer, or inter-system crossing.\cite{Ramirez2021,Yamilova2020WhatIs} However, the dominant degradation processes in NFAs include intermixing and demixing of the heterojunction components, which can be controlled to some extent by tuning crystallinity, miscibility, and the diffusion coefficient of the NFA.\cite{du2020am, Wang2021Recent, Hu2020TheRole, Alam2020Disentanglement} Structural decomposition of acceptors such as IT-4F \cite{Liu2020visible, Sudakov2020TheInterplay} occurs near to zinc oxide surfaces and under the influence of oxygen. In some cases, the device stability can be improved by increasing the molecular planarity.\cite{Sudakov2020TheInterplay} In the case of PM6:IT-4F, the degradation mechanism is the same regardless of whether devices are aged thermally in the dark or at room temperature under 1 sun illumination (without UV).\cite{du2020am}

In this work, we use a combination of experiments and simulation to investigate the mechanisms leading to degradation of devices based on the state-of-the-art PM6:Y6 material system. We identified two degradation pathways: 1.\ a $\Jsc$ loss for aging under illumination in air; 2.\ a loss of open-circuit voltage $\Voc$ and fill factor $\FF$ for aging at elevated temperatures of 85~\degreeC under nitrogen atmosphere in the dark. Investigating the more severe effects of thermal degradation in detail, we find that material properties and interface energetics remain remarkably stable. However, the devices suffer from trap state formation in the bulk. The increased charge carrier trapping causes larger non-radiative recombination losses and thus reduces $\Voc$, as well as the charge carrier mobility. Trap formation in the photoactive layer increases large initial transport resistance even further. We conclude that despite the mostly stable material properties, aging-induced trap states with their impact on non-radiative recombination and transport limitations remain the key issue to be tackled in order to improve state-of-the-art OSC stability. For organic devices to make the leap from the lab to large scale production the community must be able to better understand and control the link between chemical structure/morphological stability and the density and distribution of transport trap states. This should be the next frontier of OSC research.

\section{Results}

\begin{figure}
    \centering
    \includegraphics[width=0.46\textwidth]{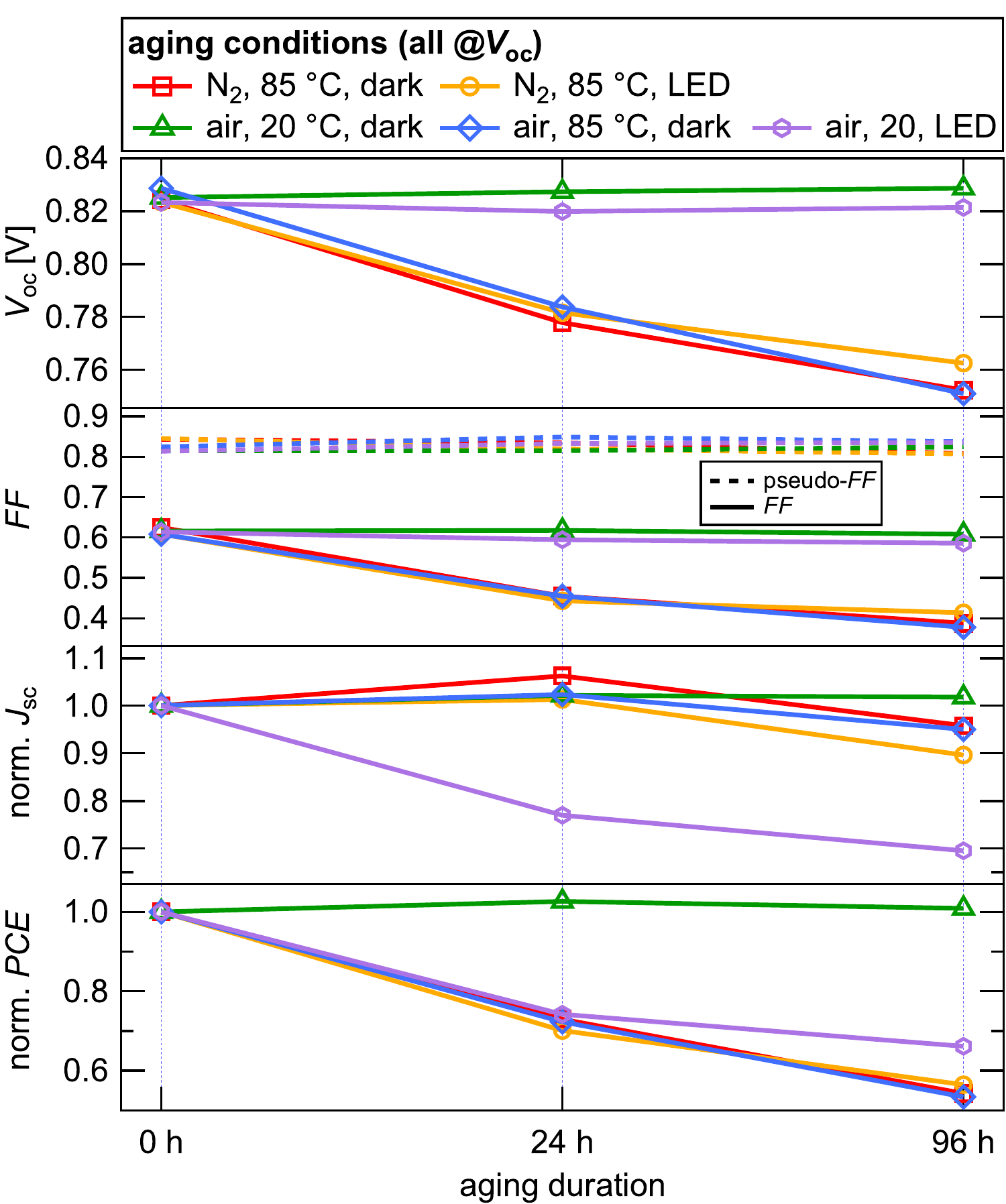}
    \caption{Solar cell parameters of devices aged under various conditions. Samples aged in air at room temperature in the dark maintain their performance over the whole aging duration of 96~hours studied here. Degradation in air at room temperature under illumination is governed by losses of $\Jsc$. Aging at elevated temperatures in air in the dark as well as in nitrogen atmosphere regardless of illumination all show losses of $\Voc$ and $FF$. The $pFF$ is stable for all aging conditions studied here.}
    \label{fig:b7panel}
\end{figure}

\begin{figure*}
    \centering
    \includegraphics[width=1.0\textwidth]{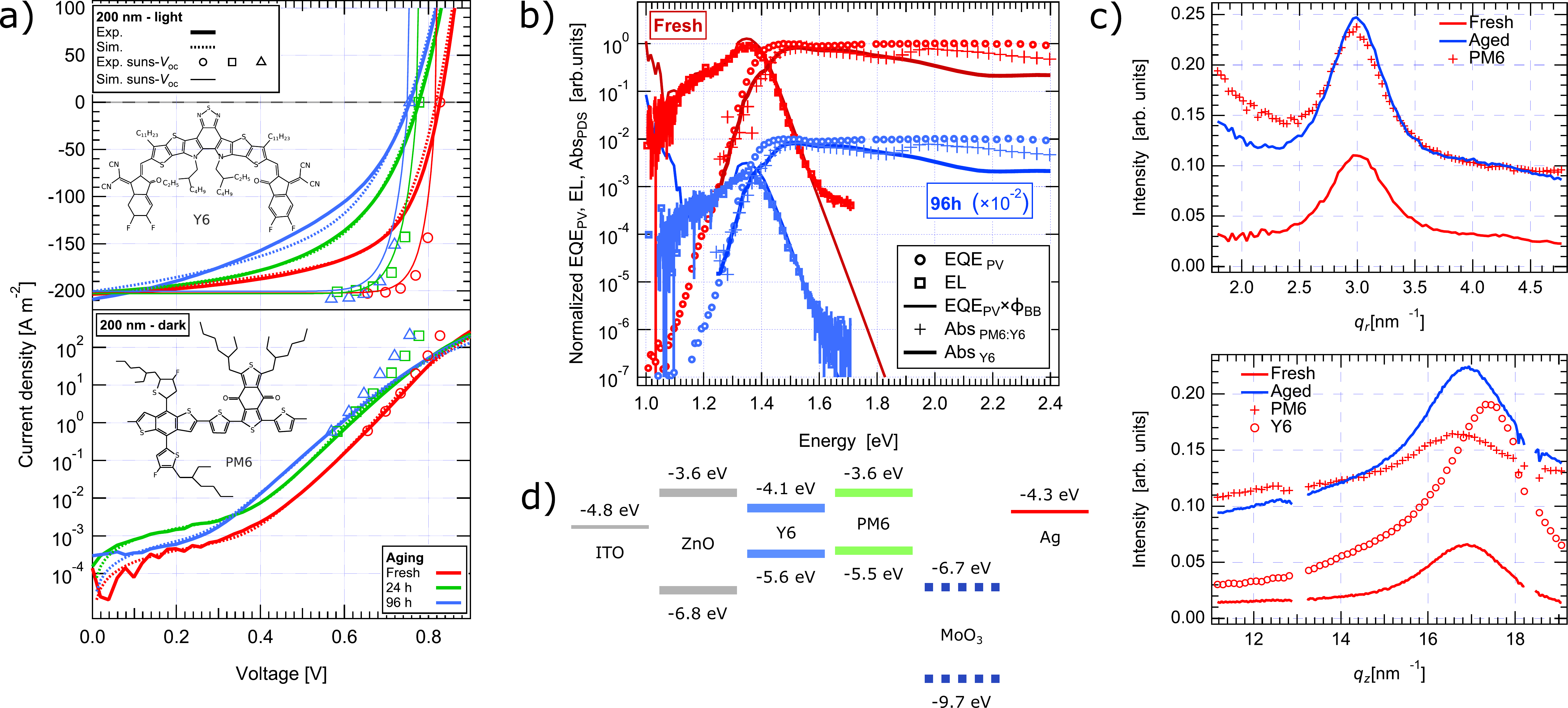}
    \caption{a) Illuminated (top) and dark (bottom) current--voltage curves of fresh and aged $200~$nm devices. For reference, the molecular structures of PM6 and Y6 are shown. b) Normalized absorption ($\mathrm{Abs}$) obtained using photothermal deflection spectroscopy, photovoltaic external quantum efficiency ($\eqepv$), and electroluminescence emission ($\el$) for fresh and 96~h samples at room temperature. Data from the aged sample was shifted down by two orders of magnitude for clarity. By comparing the external quantum efficiency $\eqepv$ multiplied by the 300~K black body spectra ($\eqepv\cdot\phi_\mathrm{BB}$) to $\el$, it can be seen that they fit very closely, suggesting through the reciprocity relation that the active layer absorption is dominated by Y6 around the 1.4~eV optical bandgap. c) Extracted horizontal (upper graph) and vertical information (lower graph) from 2D GIWAXS data of PM6:Y6/ZnO/ITO samples. The gaps seen in the vertical information correspond to missing $q$-values due to detector gaps. d) The schematic band diagram of the device.}
    \label{fig:panel1}
\end{figure*}

Several sets of glass/ITO/ZnO/PM6:Y6/MoO$_\text{x}$/Ag devices with active layer thicknesses of 200~nm were fabricated. The band diagram of the device and molecular structures are shown in Fig.~\ref{fig:panel1}. To study the effects of different aging conditions, we exposed the devices under electrical open-circuit conditions to nitrogen atmosphere and air, at room temperature and elevated temperature of 85~\degreeC as well as in the dark and under illumination by a warm-white LED. The devices were characterized when fresh and again after 24 and 96~hours of aging. The evolution of the samples' photovoltaic parameters is shown in Fig.~\ref{fig:b7panel}. We observe a remarkable stability of the devices aged in the dark in air at room temperature, as solar cell performance did not change during the aging procedure.

For a device aged in air at room temperature under illumination, its degradation features a drop of $\Jsc$ to 70~\% of its initial value after 96~hours, while $\Voc$ and $FF$ remain stable. However, device degradation for aging at elevated temperature is governed by losses of $\Voc$ and $FF$ with $\Jsc$ remaining overall stable. Under those conditions, $\Voc$ dropped from 825~mV when fresh to $\approx$780~mV after 24~hours and below 760~mV after 96~hours. The $FF$ was reduced from initially 62~\% to below 50~\% after 24~hours and further below 40~\% after 96~hours.

We have also calculated and shown the pseudo-fill factor $pFF$, which is the fill factor of the series resistance-free current--voltage characteristics determined by suns--$\Voc$ measurements. The $pFF$ is generally unaffected by charge transport and, thus, only carries information about recombination processes. Since $pFF$ is stable at 82~\% for all aging conditions studied here, we conclude that the heavily reduced $FF$ of thermally-aged devices must be dominated by processes involving charge transport limitations rather than recombination processes.

From these data we have identified two distinct pathways for degradation in the devices studied here. The first one features a loss of $\Jsc$ and only shows up for devices aged under illumination in air. Since such a $\Jsc$ loss is neither observed in devices aged under illumination in nitrogen atmosphere, nor in the dark in air, we conclude that this pathway requires both photoexcitation and oxygen. A reasonable degradation mechanism is the recombination of excited triplet excitons to the ground state by simultaneously exciting an oxygen molecule. Those oxygen radicals have been identified as the precursor of chemical reactions in a polymer:NFA solar cell leading to photobleaching and, thus, to $\Jsc$ losses.\cite{Sudakov2020TheInterplay,Ramirez2021,Liu2021Emerging,privitera2021combining} The second degradation pathway is characterized by losses of $\Voc$ and particularly $\FF$, and is observed whenever the device under test is aged thermally. Since the losses relative to the initial values are the same for aging in air and under nitrogen atmosphere, the underlying degradation mechanism does not seem to require the presence of oxygen. Apart from that, the detailed mechanism behind this degradation mechanism is unclear to date. We therefore focus our study on the mechanism behind thermally induced degradation in those solar cells.

To investigate the origin of degradation upon thermal aging in more detail, two further sets of devices with active layer thicknesses of 100~nm and 200~nm were fabricated using the automated cell preparation robot AMANDA.\cite{Du2021amanda} The devices were again characterized when fresh and after 24 and 96~hours of aging at 85~\degreeC in the dark under nitrogen atmosphere. The corresponding current--voltage curves are shown in Fig.~\ref{fig:panel1}a. 

\textbf{Charge photogeneration is not impacted by thermal aging.} 
The short circuit current $\Jsc$ remains almost unchanged upon aging (Fig.~\ref{fig:panel1}a), showing that the diffusion length of excitons -- mainly generated in pure donor or acceptor domains -- relative to the donor--acceptor domain size is still sufficient to harvest the major fraction of singlet excitons for charge photogeneration.

To study the impact of aging on charge photogeneration, we performed fast transient absorption spectroscopy on fresh and thermally aged films of PM6:Y6 blends. When excited at 800~nm, the dynamics of the Y6 exciton (band at 920~nm, Supplementary Fig.~\sref{fig:SI:TA}) that feeds the charge population saw negligible differences with aging. From this, we conclude that the rate of exciton dissociation into charges remains unaffected by thermal stress. Perdig{\'{o}}n-Toro et al.\cite{perdigon2020barrierless}\ reported an electric field-independent charge photogeneration in fresh PM6:Y6 solar cells. Combined with our findings, it is unlikely that charge generation is causing the decrease in fill factor (Fig.~\ref{fig:panel1}a) associated with degradation of the devices over time.

Fig.~\ref{fig:panel1}b shows the sub-bandgap energy landscape as measured using photothermal deflection spectroscopy (absorption), external quantum efficiency ($\eqepv$) and electroluminescence ($\el$) spectroscopy. We observe that both fresh and 96~h aged solar cells have almost identical spectra, which is consistent with the ellipsometry results (see Supplementary Fig.~\sref{fig:SI:ellipsometry}). The absorption of Y6, also shown in Fig.~\ref{fig:panel1}b, is very similar to that of PM6:Y6. From this resemblance we conclude that the optical bandgap and the distribution of tail states in the blend are dominated by Y6. The tail states can be approximated by an exponential distribution with an Urbach energy of 21.5~meV for 200~nm thick devices (24.5~meV for 100~nm). Aging does not affect the distribution width (see Supplementary Fig.~\sref{fig:SI:UrbachEnergies_vs_Thickness_Aging}). The electroluminescence emission at room temperature closely follows the calculated electro-optical reciprocity ($\eqepv\cdot\phi_\mathrm{BB}$) with the 300~K black body emission spectra $\phi_\mathrm{BB}$. As the optical bandgap around 1.4~eV is dominated by Y6 absorption, the reciprocity suggests that the acceptor also dominates the emission.\cite{perdigon2021excitons}

\textbf{Morphology: paracrystalline disorder increases for $\pi$--$\pi$-stacking of polymer.}

While the optical characteristics are not changing much upon degradation, the current--voltage characteristics (Fig.~\ref{fig:panel1}a) show a reduction of the fill factor and the open-circuit voltage. To better understand if changes to the active layer morphology upon aging can be responsible, we performed grazing incidence wide angle X-ray scattering (GIWAXS) on samples of PM6:Y6 spin coated onto both silicon and ITO/ZnO substrates. We measured the films before and after 96~h of aging, the results on ITO/ZnO are shown in Fig.~\ref{fig:panel1}c. For the PM6:Y6 films on Si (Supplementary Fig.~\sref{fig:SI:herzig1}), measurements reveal a significant increase in order upon aging, particularly for Y6. In the small $q$ region, peaks corresponding to the characteristic nanostructure of long range ordered Y6\cite{Zhu2020} are observed initially weakly, but very strongly after aging. In contrast to the PM6:Y6 blends prepared on silicon substrates, those prepared on ITO coated with ZnO show only a weak Y6 nanostructure and in comparison more subtle aging effects (Fig.~\ref{fig:panel1}c and Supplementary Fig.~\sref{fig:SI:herzig2}). Thus, the ZnO layer seems to reduce aggregation and strong morphological changes within the blend samples upon aging. To understand the relationship to the device characteristics better, we focus on a more detailed examination of the PM6 peak ($q_r$ = 3.0~nm\textsuperscript{-1}) in the ZnO coated sample. This peak has Lorentzian shape with a paracrystalline disorder parameter\cite{peng2000arxiv} $g$ that minorly reduces its value from fresh to aged blend sample, while the intensity of the peak increases by almost 80\%. This implies that the order within lamellar stacks of PM6 increases only slightly, but that a larger fraction of PM6 exhibits lamellar stacking with aging (Fig.~\ref{fig:panel1}c). For Y6, no significant changes can be identified from the data.

On the other hand, the examination of the $\pi$--$\pi$-stacking distance of the blend samples on ZnO in the vertical direction reveals that the PM6 initially has a higher stacking order to start with. During aging, the paracrystallinity disorder parameter $g$ in the $\pi$--$\pi$-stacking of PM6 increases along with higher peak intensity ($>$50\%), indicating more material is contributing to the $\pi$--$\pi$-stacking signal with the peak shape of the 96~h aged sample approaching towards that of the neat material ($g_\textrm{blend,fresh}=(13.5\pm 0.3$\%, $g_\textrm{blend,aged}=(16.3\pm0.5)$\%, $g_\textrm{neat,PM6}=(17.4\pm 0.3)$\%). The loss in order might be surprising, but can be explained with a competition between different stacking mechanisms.\cite{peng2000arxiv} For PM6 blended with Y6, the increase in disorder of the $\pi$--$\pi$-stacking in unison with the increase in the amount of lamellar stacking implies that the lamellar stacking is enhanced at the cost of the $\pi$--$\pi$-stacking quality. We expect that these changes have an impact on the energetic disorder and can be the origin for the negative impact of aging on the charge transport properties, which are discussed further below.

\begin{figure*}[tb]
	\centering
	\includegraphics[width=\textwidth]{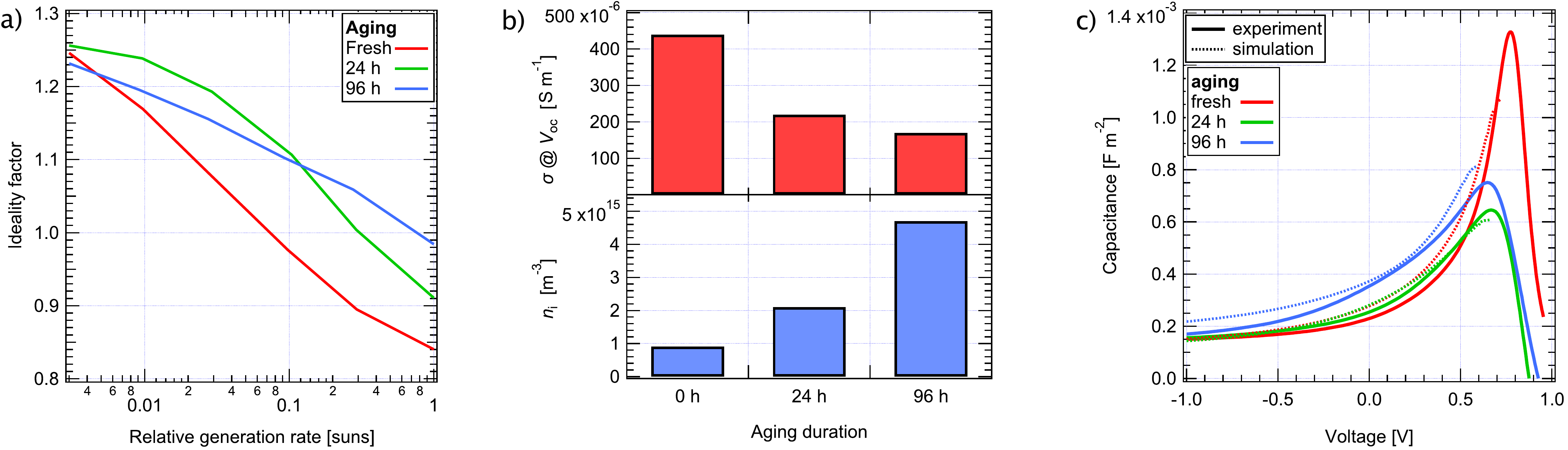}
	\caption{a) Ideality factor of fresh and aged devices for illumination intensities ranging from 0.003~suns to 1~sun. b) The electrical conductivity $\sigma$ near open-circuit conditions and the estimated intrinsic charge carrier density $n_\mathrm{i}$, for various aging durations. c) Experimental and simulated $CV$ curves under 1~sun equivalent illumination.}
	\label{fig:panel2}
\end{figure*}

\textbf{Which mechanism limits the open-circuit voltage?}
We performed the suns--$\Voc$ measurements\cite{sinton2000, Schiefer2014determination} to identify the dominant recombination mechanisms at various light intensities (cf.\ Fig.~\ref{fig:panel1}a). The corresponding ideality factors $\nid$ in Fig.~\ref{fig:panel2}a indicate a transition from trap-assisted bulk recombination to surface recombination at higher illumination intensities. The ideality factor for a given generation rate $G$ was determined by intensity-modulated photovoltage spectroscopy (IMVS) as
\begin{equation}
    \nid=\frac{e}{kT}\frac{\operatorname{Re}\left(\mathrm{IMVS}\right)}{\ln\left(\frac{G+G_\mathrm{AC}}{G}\right)}\quad .
    \label{eq:nid_IMVS}
\end{equation}
Here, $G_\mathrm{AC}$ is the generation rate change due to the modulated light intensity amplitude, $e$ is the elementary charge, $k$ Boltzmann's constant, and $T$ is the absolute temperature.

The values of $\nid$ up to 0.3~suns are between 1.1 and 1.2  independent of the aging duration. For higher illumination intensities, $\nid$ first approaches and then drops below unity. Therefore, the open-circuit voltage under 1~sun illumination is limited by surface recombination. As we do not observe s-shaped $JV$ curves, we do not refer to an extraction-rate limited photocurrent,\cite{wagenpfahl2010} but an open-circuit voltage limited by the work function difference of the electrodes.\cite{rauh2011relation,Solak2016}

To explore whether the energetic alignment at the interface between the organic active layer and the MoO$_\text{x}$/Ag electrode is changing upon aging, we performed ultraviolet photoemission spectroscopy (UPS) measurements on fresh and aged devices. We find that, within the experimental error of UPS ($\approx 0.1$~eV), neither the work function, nor the ionization potential of the surface of the active layer changes upon thermal stress for 96~h (see Supplementary Fig.~\sref{fig:SI:UPS}). These measurements confirm that the energetic alignment -- and thus, the built-in potential of the device -- is unaffected by aging.

From measurements of the relative electroluminescence quantum yield (see Supplementary Fig.~\sref{fig:SI:EQEled_dVoc}) it is evident, however, that upon aging $\Voc$ is increasingly limited by non-radiative recombination losses. $\Delta V_\mathrm{oc,nonrad}$, calculated from the relative difference of integrated $\el$ spectra of aged samples compared to fresh ones, increases with aging duration by $\approx 80$~mV after 24~h and $\approx 100$~mV after 96~h. These relative changes reflect the trend of $\Voc$ decreasing from $\approx 830$~mV to $\approx 755$~mV during aging.
The dark saturation current $J_0$ increases accordingly, confirming that while surface recombination limits the open-circuit voltage, the effect of aging on $\Voc$ manifests itself mainly by the higher degree of non-radiative, trap-assisted recombination.

\textbf{The effective bandgap and singlet exciton repopulation.}
Another perspective on the open-circuit voltage is to consider it as the effective bandgap reduced by radiative and non-radiative recombination losses. By extrapolating the temperature-dependent open-circuit voltage to 0~K\cite{rauh2011relation} in Supplementary Figure~\sref{fig:SI:VocGT_Analysis}, we determine the apparent effective bandgap to $1.1~$eV. We find no significant effect of active layer thickness or aging. 

The significant difference of the apparent effective bandgap to the optical bandgap $E_\mathrm{opt}=1.38~$eV,\cite{perdigon2021excitons} observed for fresh and aged devices, cannot be explained by the small energy loss of $\approx 50~$meV during the charge transfer.\cite{karki2019understanding} 
It was reported recently that the electroluminescence quantum yield $\mathrm{QY_{EL}}$ in PM6:Y6 blends exhibits a significant temperature dependence from repopulation of singlet exciton states $S_1$ in Y6 via an energy barrier of $\Delta E_\mathrm{S_1\text{-}CS}\approx 0.12~$eV.\cite{perdigon2021excitons} Recombination from the repopulated $S_1$ states causes additional losses to the quasi-Fermi-level splitting of the CT state, and should result in a lower $\Voc$. With this in mind, we suggest that the extrapolated $\Voc$ at 0~K underestimates the effective bandgap, and has to be corrected by $\Delta E_\mathrm{S_1\text{-}CS}$ (see Supplementary Information for further details), yielding $E_\mathrm{g,eff}=1.22$~eV. However, as we find thermal aging affects neither apparent nor corrected effective bandgap, we expect the radiative fraction of the $\Voc$ losses to be unchanged by the thermal aging.

\textbf{The fill factor is mainly limited by the transport resistance rather than non-geminate recombination.} 
The $JV$ curves under illumination are shown in Fig.~\ref{fig:panel1}a, along with the corresponding $JV$ curves without the negative influence of the transport resistance.\cite{PYSCH20071698} The latter were calculated by subtracting $\Jsc$ at 1~sun from the suns--$\Voc$ data, and show the impact of non-geminate recombination without the influence of other losses. The light $JV$ curves are, in contrast, additionally limited by the transport resistance, which already reduces the fill factor of the fresh devices more strongly than non-geminate recombination. In other words, the voltage difference $\Delta V$ between the light $JV$ curve and the shifted suns--$\Voc$ curve at a common current density $J$ increases upon aging. 

$\Delta V$ can be understood as the voltage loss due to the transport resistance $R_\mathrm{tr}$ of the active layer.\cite{PYSCH20071698,mackel2018determination,Schiefer2014determination} As the transport resistance originates from a low conductivity, that means that e.g.\ a low charge carrier mobility can lead to a decoupling of the externally applied voltage and the internal one of the active layer: $\Delta V = V_\mathrm{applied} - V_\mathrm{internal}$. The effective single-carrier conductivity
\begin{equation}
	\sigma=\frac{L}{2R_\mathrm{tr}}=\frac{JL}{2\Delta V(J)} \quad ,
	\label{eq:conductivity}
\end{equation}
with the active layer thickness $L$, is shown in Fig.~\ref{fig:panel2}b for voltages close to $\Voc$. The fresh device has a conductivity of $4.0\cdot10^{-4}~$S~m$^{-1}$, which is reduced to $2.2\cdot10^{-4}~$S~m$^{-1}$ and $1.7\cdot10^{-4}~$S~m$^{-1}$ after 24~h and 96~h of aging, respectively. The analog results for the 100~nm thick active layer devices and the weaker fill factor losses are consistent with $\Delta V \propto L$ in Eqn.~\eqref{eq:conductivity}. Thus, the dominant loss in fill factor for fresh and aged devices is due to the conductivity-limited transport resistance. This is in accordance with our interpretation of an increased density of trap states in the bulk of the active layer.

\textbf{Evidence for aging-induced trap states.}
In order to understand the experimental results in a holistic fashion, we use a drift--diffusion model which describes carrier capture/escape using a time domain Shockley-Read-Hall formalism. The general purpose photovoltaic device model (gpvdm) was described elsewhere in detail.\cite{mackenzie2012extracting} For this work, we extended the model to the frequency domain, and the model was self-consistently fit simultaneously to the dark $JV$ curves (Fig.~\ref{fig:panel1}a bottom), illuminated $JV$ curves (Fig.~\ref{fig:panel1}a top), $CV$ data (Fig.~\ref{fig:panel2}c), suns--$\Voc$, $J_{sc}$ vs.\ light intensity, and the real/imaginary parts of the impedance spectroscopy data (see Supplementary Information). This allowed us to extract a set of material parameters as a function of aging (see Supplementary Tables~\sref{SI:tab:sim_E6} and~\sref{SI:tab:sim_E10}).

The extracted material parameters suggest that, as the devices age, the density of trap states in the bulk heterojunction gradually rises (from about $10^{24}$~m$^{-3}$ to about $10^{26}$~m$^{-3}$), and the carrier capture cross sections gradually change, suggesting some level of microscopic reordering while the \emph{free} carrier mobility remained constant. These changes are accompanied by a reduced \emph{effective} charge carrier mobility:\cite{goehler2018} under open-circuit conditions at 1~sun, $\mu$ decreases from approximately $5\cdot10^{-8}~\mathrm{m^2V^{-1}s^{-1}}$ to $1$--$2\cdot10^{-8}~\mathrm{m^2V^{-1}s^{-1}}$ after 96~h of aging (Supplementary Fig.~\sref{fig:SI:n_and_taun}). The correspondingly decreased simulated conductivity is consistent with the experimentally determined values of $\sigma$,\cite{Schiefer2014determination} calculated by Eqn.~\eqref{eq:conductivity} and shown in Fig.~\ref{fig:panel2}b, which decrease with aging duration for all thicknesses to approximately half the respective initial value after 96~h.

Fig.~\ref{fig:panel2}c shows the experimental and simulated capacitance--voltage curves under 1~sun illumination, from negative voltage to just over $\Voc$. At large negative voltages the electric field between the contacts sweeps carriers out of the device quickly, thus the capacitance tends towards the geometric capacitance. In this region the values of capacitance are very close to one another, suggesting no significant change in the structure of the device upon aging. As the voltage increases from negative to positive, the field between the contacts drops, diffusion currents start to dominate drift currents, and charge carriers start to interact with trap states. Examining the range -1--0.5~V, it can be seen that the capacitance of the 96~h device increases most rapidly as voltage is swept from negative to positive, followed by the 24~h aged device and then the fresh device. This suggests that the trapping density increases as a function of aging duration.\cite{Wang2021Recent}

The emergence of aging-induced energetic traps is further indicated by an increased intrinsic carrier concentration $n_\mathrm{i}$ during aging. As traps are described by a distribution of energetic states extending into the (effective) bandgap of the active-layer material, an overall increased trap density leads to higher $n_\mathrm{i}$ due to the stronger overlap with the Fermi distribution. We evaluated the experimental $CV$ curves to determine the effective charge carrier density $n$ as well as an estimate for the intrinsic charge carrier density $n_\mathrm{i}$. $n$ is obtained from the chemical capacitance $C_\mu\left(V\right) = C(V) - \text{geometric capacitance}$, using the equation\cite{Elliott_2014,Vollbrecht2020oncharge}
\begin{equation}
    n\left(\Voc\right) = n\left(V_\mathrm{sat}\right)+\frac{1}{eL}\int_{V_\mathrm{sat}}^{\Voc} C_\mu\left(V\right)\d V\quad.
	\label{eq_n}
\end{equation}
$V_\textrm{sat}$ is a voltage in the reverse-bias regime, where non-geminate recombination is heavily suppressed. With $n\left(V_\mathrm{oc}\right)$, we achieve an estimate for $n_\mathrm{i}$ of the active layer blend,\cite{Schiefer2014determination}
\begin{equation}
    n_\mathrm{i} = n\left(V_\mathrm{oc}\right)\exp\left(-\frac{eV_\mathrm{oc}}{2kT}\right) \quad .
    \label{eq_ni}
\end{equation}
The resulting intrinsic carrier concentration $n_\mathrm{i}$, shown in Fig.~\ref{fig:panel2}b, increases from $1\cdot10^{15}~\mathrm{m}^{-3}$ for the fresh device to about $5\cdot10^{15}~\mathrm{m}^{-3}$ after 96~h of aging (and $\approx1\cdot10^{15}~\mathrm{m}^{-3}$ to $\approx4\cdot10^{15}~\mathrm{m}^{-3}$ for the 100~nm device). We interpret this increase of $n_\mathrm{i}$ as an indication for an aging-induced increasing trap density extending more deeply into the effective bandgap, which is consistent with the decreasing effective charge carrier mobility and the reduced conductivity.


\section{Discussion} 

We found two degradation pathways in inverted solar cells with a bulk heterojunction active layer consisting of PM6:Y6. The first pathway requires the presence of both oxygen and photoexcitation and features a loss of $\Jsc$. This can be potentially explained by chemical reactions of the active layer components with singlet oxygen, formed by recombination of triplet excitons. The second degradation pathway is characterized by simultaneous losses of $\Voc$ and $FF$ and is enabled by elevated temperature regardless of atmospheric or illumination conditions. 

We have focused our investigation on the less well-known effects of accelerated thermal aging in the dark. A variety of electrical and optical experimental methods as well as drift--diffusion simulations were used to achieve a more general understanding of processes leading to degradation of this state-of-the-art NFA-based photovoltaic system. We find that, while absorption and the effective bandgap remain unaffected, the aging-induced performance reduction in the studied blend is caused by an increasing trap density in the bulk of the active layer. We believe that the reduced $\pi$--$\pi$-stacking quality we observe is responsible for this change. The higher trap density leads to increased non-radiative recombination and transport resistance losses, ultimately lowering the solar cell performance by a reduction of both, open-circuit voltage and fill factor. Furthermore, we find the trap-induced transport resistance to be the major reason for the drop in fill factor. We conclude that despite the mostly stable material properties, the formation of trap states upon thermal degradation with their impact on non-radiative recombination and transport limitations remain the key issue to be addressed for improving the state-of-the-art OSC stability: This is the next frontier of OSC research.


\section{Methods}

\textbf{Materials.}
PM6 and Y6 were purchased from Solarmer Materials Inc (Beijing, China) and Derthon Optoelectronic Materials Science Technology Co LTD (Shenzhen, China) and used as received. ZnO nanoparticles (N10) were received from Avantama AG.

\textbf{Solar Cell Devices.}
Inverted devices based on the PM6:Y6 material system were fabricated in a device architecture of glass/ITO/ZnO/PM6:Y6/MoO$_\text{x}$/Ag. ITO-coated glass substrates were successively cleaned with deionized water and isopropanol about 15 min, respectively. The ZnO nanoparticles solution was spin-coated onto ITO substrate at 3000 rpm for 30 s followed by thermal annealing at 200~\degreeC for 30 min. The substrates were then transferred into a nitrogen-filled glove box. The PM6:Y6 (1:1.2) blends were dissolved in chloroform (CF) at a total weight concentration of 14.3 mg/mL and 22 mg/mL to fabricate 100 nm and 200 nm thick active layers, respectively. Before spin-coating of the active layer, 0.5 vol.-$\%$ 1-chloronaphthalene (CN) was added into the PM6:Y6 solution as solvent additive. The active layers were spin-coated onto the substrates and thermal annealed at 100~\degreeC for 10 min. Subsequently, 10~nm MoO$_\text{x}$ and 100~nm Ag were thermally evaporated with a shadow mask on top of the active layer as hole-transport layer and electrode, respectively. The thicknesses were confirmed by capacitance spectroscopy at reverse bias. 

\textbf{Aging.}
All devices studied here have been aged under open-circuit conditions and characterized when fresh and after 24 and 96~hours of aging. For the main part of this manuscript, thermal aging was applied in the dark and in a nitrogen-filled glovebox to study the degradation pathway characterized by losses of $FF$ and $\Voc$. For a broader overview, various different aging conditions have been applied to a number of samples. Namely, conditions have been:
\begin{itemize}
    \item
        nitrogen atmosphere, 85~\degreeC, dark
    \item
        nitrogen atmosphere, 85~\degreeC, illuminated
    \item 
        nitrogen atmosphere, room temperature ($\approx20$~\degreeC), illuminated
    \item 
        air, room temperature, dark
    \item
        air, room temperature, illuminated
    \item
        air, 85~\degreeC, dark
\end{itemize}
For illumination we used an array of warm-white LEDs without UV wavelengths. The illumination intensity was insufficient for reaching a 1~sun-equivalent intensity. To estimate the illumination intensity, the short-circuit current of a silicon photodiode under LED illumination relative to its short-circuit current under AM1.5 illumination was calculated. That way, the LED illumination intensity was estimated to approximately $0.5$~sun-equivalent.

\textbf{Grazing Incidence Wide Angle X-Ray Scattering (GIWAXS).} 
GIWAXS experiments were conducted at the beamline 7.3.3 at the Advanced Light Source at Lawrence Berkeley National Lab (Berkeley, USA)\cite{hexemer2010}. The samples were illuminated with 10~keV radiation ($\lambda = 1.24$~\AA) at an incident angle ($\alpha_{i}$) of 0.18$^{\circ}$. The beam size was 300~$\mu$ m (height) $\times$ 700~$\mu$ m (width). The scattering signal was captured on a Pilatus 2M (172~$\mu$ m pixel size, file format EDF, 1475x1679 pixels) located 275~mm from the sample. Acquisition times were 10~s for each frame. q-conversion of 2D pixel data and 1D data reduction was done by XSACT software from Xenocs. To track scattering peak parameters the 1D intensity profiles were fitted with Gaussians and Lorentz-functions using a Trust Region Reflective Least-Squares algorithm and analyzed according to~\cite{peng2000arxiv}. The 2D data is shown in Fig.~\sref{fig:SI:herzig1} and Fig.~\sref{fig:SI:herzig2}.

\textbf{Variable angle spectroscopic ellipsometry.}
Spectroscopic ellipsometry measurements were performed using an M2000 ellipsometer from J. A. Woollam Co. The spectra were recorded with a spectral range of 0.735 to 5.042~eV at variable angles of incidence from 45\textdegree to 75\textdegree in the step of 5\textdegree. The data modeling and analysis were performed on a simulation and spectrum modeling software CompleteEASE from J.A.\ Woollam Co. A layer-by-layer optical, analogous to the physical layer structure was designed. For high reliance on the results, a multi-sample analysis (MSA) approach was used to determine the optical constants of the unknown layer. The MSA allows multiple spectra to be fitted simultaneously with one or more common fit parameters to all spectra while keeping some as independent fit parameters that are allowed to be varied for individual spectra.

\textbf{Photothermal deflection spectroscopy (PDS).} 
PDS measurements were performed following the previously developed procedure.\cite{Becker_Koch_2019} In short, the organic films on quartz substrate were mounted in the signal enhancing liquid (Fluorinert FC-770) filled quartz cuvette inside a N\textsubscript{2} filled glovebox. Then, the samples were excited using a tunable, chopped, monochromatic light source (150~W xenon short arc lamp with a Cornerstone monochromator) and probed using a laser beam (635~nm diode laser, Thorlabs) propagating parallel to the surface of the sample. The heat generated through the absorption of light changes the refractive index of the Fluorinert liquid, resulting in the deflection of the laser beam. This deflection was measured using a position sensitive-detector (Thorlabs, PDP90A) and a lock-in ampliﬁer (Amatec SR7230) and is directly correlated to the absorption of the film.

\textbf{Ultraviolet photoemission spectroscopy (UPS).} 
UPS measurements were performed on photovoltaic devices that were heat stressed for 0, 25, 50 and 100~h prior to the removal of the Ag electrode in order to expose the interface with the organic active layer. The UPS spectra were collected using a PHOIBOS 100 analyzer system (Specs, Germany) at a base pressure of $2\times 10^2$~mbar using a He I excitation line (21.22~eV).

\textbf{Current--voltage characterization.} 
A Keithley~236 SMU was used for voltage application and current measurement. AM1.5 illumination was provided by a Wavelabs LS-2 solar simulator. No  aperture was used. The illumination was kept switched on for two seconds per measurement to prevent the sample temperature from increasing.

\textbf{Impedance spectroscopy, intensity-modulated spectroscopy.} Modulated and continuous illumination was provided by an Omicron A350 diode laser with a center wavelength of 515~nm. A Zurich Instruments MFLI lock-in amplifier with MF-IA, MF-MD, and MF-5FM options was used to measure sample current and voltage as well as providing voltage to modulate the laser. The illumination intensity was varied using neutral density filters mounted in a Thorlabs motorized filter wheel FW102C combined with a continuously variable neutral density filter wheel. For IMPS and IMVS measurements, the amplitude of modulated illumination was chosen to be 10\% of the bias illumination intensity to ensure small-signal excitation. Laser calibration was performed using a Newport 818-BB-21 biased silicon photodetector. Capacitance--voltage measurements were performed with a voltage-modulation frequency of 50~kHz and an amplitude of 20~mV.

\textbf{Sensitive external quantum efficiency (EQE).}
$\eqepv$ were measured by photocurrent spectroscopy, excited by light from a $150~$W quartz-tungsten-halogen lamp passing a LOT-QD MSH-D300 double monochromator and additional optical long pass filters to reduce stray light. The incoming light was mechanically chopped at a frequency of $419~$Hz and a small part was directed to a Hamamatsu K1718-B two-color photodiode for reference. The solar cell photocurrents were pre-amplified by a variable Zurich Instruments HF2TA current--voltage amplifier and the resulting voltages again amplified and detected by a Zurich Instruments HF2LI lock-in amplifier.

\textbf{Electroluminescence (EL) spectroscopy.}
EL spectra were recorded with liquid-nitrogen cooled Princeton Instruments Spec10:100 silicon and PyLoN-IR cameras in combination with a Acton SpectraPro500i spectrograph. Constant driving currents of $96~\mathrm{mA~cm^{-2}}$ were applied to the solar cells from a Keithley~2635a SMU.

\textbf{Open-circuit voltage spectroscopy.}
For recording temperature dependent $\Voc$, the solar cells were held in a custom-built closed-cycle contact gas cryostat (CryoVac) and excited by an Nd:YAG continuous-wave laser (Millenia Pro) set to an output power of $1~$W. Different illumination intensities in a range of 9 orders of magnitude were realized by the laser passing two neutral density filter wheels (Thorlabs) with discrete optical density steps. A mechanical shutter was used to reduce irradiation between measurement steps in order to avoid light induced heating of the sample. Photovoltages were recorded by a Keithley~2635a SMU.

\textbf{Suns--$\Voc$ measurements.}
Sample photoexcitation was provided by an Omikron LDM A350 continuous wave laser operating at a wavelength of 515 nm. Various illumination intensities have been realised by motor-driven neutral-density filter wheels by Thorlabs and Standa, respectively. $\Voc$ and $\Jsc$ have been measured with a Keithley~2634b SMU. Suns--$\Voc$ data were calculated from measured $\Jsc\left(\Voc\right)$ data by substracting $\Jsc$ under 1~sun-equivalent illumination.

\textbf{Transient absorption (TA) spectroscopy.}
TA was carried out similarly to previously described pump-probe experimental schemes. The particularities of our setup are as follows. A Ti:sapphire laser system (Coherent Astrella-F-1K) with a 1~kHz repetition rate and a fundamental output centered at 800~nm was used as the pulsed excitation source. The samples were excited at the fundamental wavelength, with a beam diameter of 1885~\textmu{}m and a fluence of $< 20$~\textmu{}J\,cm$^{-2}$ to assure that excitation density remained within the linear regime. The pump pulses were modulated by a chopper to have half the frequency of the amplifier, 500~Hz. 
The broadband white light for the probe was generated by passing a 1600~nm beam (generated in a home-built optical parametric amplifier) through a 5~mm YAG crystal and had a beam diameter of 225~\textmu{}m at the sample. The spectra were recorded with a Hamamatsu InGaAs Camera (G11608-512DA) . The pump and probe beams were set to have magic angle relative polarizations, and they were spatially overlapped at the sample, and temporally delayed with a translation stage.

\begin{acknowledgments}
C.D.\ and his group thank the Deutsche Forschungsgemeinschaft (DFG) for funding this work (projects Photogen (no.~362992821) and TEMET NOSCE (no.~279635873)). C.Gr.\ and E.M.H.\ acknowledge support by Deutsche Forschungsgemeinschaft (DFG) through TUM International Graduate School of Science and Engineering (IGSSE) and the Bavarian State Ministry of Science, Research and the Arts through the Collaborative Research Network \enquote{Solar Technologies go Hybrid}. Portions of this research were carried out at beamline~7.3.3 of the Advanced Light Source, which is supported by the Director of the Office of Science, Office of Basic Energy Sciences, of the U.S. Department of Energy under Contract No.~DE-AC02-05CH11231.
\end{acknowledgments}

\appendix

\section*{Corresponding Author}
Email: deibel@physik.tu-chemnitz.de

\section*{Author contributions:}\ C.D.\ conceptualized the study. X.D.\ and L.N.\ fabricated the photovoltaic devices and characterized their photovoltaic performance under the supervision of T.H., N.L.\ and C.J.B. C.W.\ performed the electrical steady state and frequency domain characterization, supported by S.A., under supervision of C.D. R.C.I.M.\ performed the device simulations. C.Gr.\ and C.Z.\ performed the morphology characterization under supervision of E.M.H. D.B.-K. and Y.J.H. performed the PDS and UPS measurements, respectively, under the supervision of Y.V. K.M.Y.\ performed the transient absorption measurements under the supervision of N.B. I.M.\ performed the ellipsometry measurements under supervision of D.R.T.Z. C.W.\ wrote the manuscript with input and revisions mainly by C.D.\ and R.C.I.M. All authors contributed by discussion to the manuscript.



%

\end{document}